\documentstyle[11pt,newpasp,twoside,epsf]{article}
\def\edcomment#1{\iffalse\marginpar{\raggedright\sl#1\/}\else\relax\fi}
\reversemarginpar

\begin{document}
\title{Arecibo Measurements of Pulsar--White Dwarf Binaries: 
Evidence for Heavy Neutron Stars}

\author{David J. Nice \& Eric M. Splaver}
\affil{Physics Department, Princeton University, 
Princeton, NJ 08544 USA}
\author{Ingrid H. Stairs}
\affil{Dept. of Phys. \& Astron., U.\,B.\,C., Vancouver BC V6T 1Z1 Canada}


\begin{abstract}
We summarize constraints on neutron star masses from
ongoing timing observations of pulsar--white dwarf binaries 
at Arecibo.  The trend is toward pulsar masses 
larger than the canonical value of 1.35\,M$_\odot$.
\end{abstract}


\noindent
We have measured relativistic phenomena (orbital precession,
Shapiro delay and/or orbital decay) in Arecibo
pulsar timing measurements of several pulsar--white
dwarf binary systems.  
These measurements constrain the pulsar mass,
secondary star mass, and orbit inclination.
This observing program has been described elsewhere
(e.g., Nice et al. 2003, 2004).
Here we give a snapshot of our present 
constraints on stellar masses and orbit inclination (Figure 1, Table 1).
These results include some work done after the conference.

The most compelling case for a high radio pulsar mass is PSR\,J0751+1807.
Its orbital period is changing at a rate of
$\dot{P_b}=(6.2\pm 0.8)\times10^{-14}$
due to the emission of gravitational radiation.  Shapiro
delay is also marginally detected.  The inferred pulsar mass,
$2.1^{+0.4}_{-0.5}\,{\rm M}_\odot$ (95\% confidence), is 
substantially higher than the canonical
value of 1.35\,M$_\odot$.  Details will be given in
Nice et al. (2005).

\acknowledgements

We thank the following for collaboration and/or for supplying
pulsar data: Z. Arzoumanian,
D. C. Backer, F. Camilo, J. M. Cordes, O. Lohmer, A. G. Lyne, and
M. Kramer.  This work was supported by NSF grant 0206205 and by
an NSERC UFA and Discovery grant.  The NAIC/Arecibo
Observatory is
operated by Cornell University under a cooperative agreement with the NSF.

\begin{table}
\caption{Mass Measurements from Timing Analysis}
\begin{tabular}{r@{}lccccc}
\tableline
\multicolumn{2}{c}{Pulsar} & Orb. Per.& Ecc.          & Pulsar Mass  & Secondary Mass & Ref. \\
             &             & (days)   &               &  \multicolumn{2}{c}{(M$_{\sun}$, 95\% conf.)} & \\
\tableline
\rule[13pt]{0pt}{0pt}
J&0621+1002 & \phantom{0}8.32 & 0.002\,457 & $1.7^{+0.6}_{-0.6}$  & $0.97^{+0.43}_{-0.24}$ & 1\\[3pt]
J&0751+1807 & \phantom{0}0.26 & 0.000\,003 & $2.1^{+0.4}_{-0.5}$  & $0.19^{+0.03}_{-0.03}$ & 2\\[3pt]
J&1713+0747 &           67.83 & 0.000\,075 & $1.3^{+0.4}_{-0.3}$  & $0.28^{+0.06}_{-0.04}$ & 3\\[3pt]
B&1855+09   &           12.33 & 0.000\,022 & $1.6^{+0.2}_{-0.2}$  & $0.27^{+0.02}_{-0.03}$ & 4\\[3pt]
\tableline
\tableline
\multicolumn{7}{l}{References:\ (1) Splaver et al. 2002, (2) Nice et al. 2005,} \\
\multicolumn{7}{l}{%
\newlength\djnlength
\settowidth\djnlength{References:}
\hspace*{\djnlength}(3) Splaver et al. 2005, (4) Splaver 2004.}
\end{tabular}
\end{table}

\begin{figure}[t]
\plotfiddle{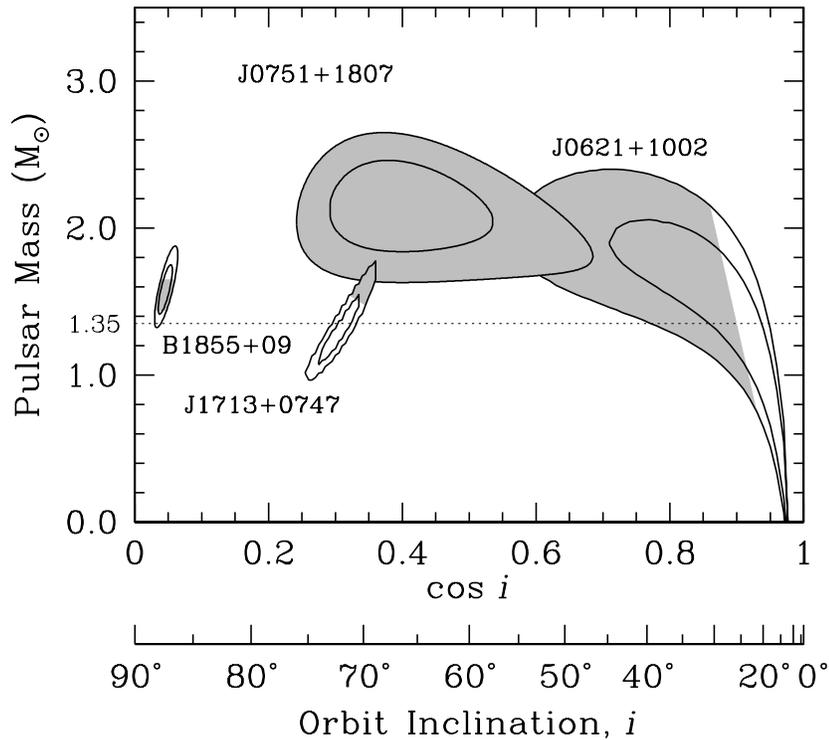}{4.6in}{0}{60}{60}{-180}{-12}
\caption{Constraints on pulsar masses and
orbit inclinations.  
Contours indicate 68\% and 95\% confidence 
regions.
Shaded areas denote theoretical limits from the
orbital period--core mass relation (B1855+09 and J1713+0747)
and from the assumption that the secondary
is a white dwarf (J0621+1002).
}
\end{figure}

\end{document}